\documentclass[aps,twocolumn,prl,superscriptaddress,amsmath,amssymb]{revtex4-1} 
\usepackage{dcolumn}
\usepackage{bm}
\usepackage{amsmath}
\usepackage{txfonts}
\usepackage[T1]{fontenc}
\usepackage{xspace}
\usepackage{ulem}
\usepackage{comment}
\newcommand{\bras}[1]{\langle#1\rvert}
\newcommand{\kets}[1]{\lvert#1\rangle}
\newcommand{\brasd}[1]{\langle\!\langle#1\rvert}
\newcommand{\ketsd}[1]{\lvert#1\rangle\!\rangle}

\newcommand{\mean}[1]{\left<#1\right>}
\newcommand{\means}[1]{\langle#1\rangle}

\ifx\pdfoutput\undefined
\usepackage[dvipdfmx]{graphicx}
\usepackage[dvipdfmx]{hyperref}
\usepackage[dvipdfmx]{color}
\usepackage[dvipdfmx]{xcolor}
\else
\usepackage{graphicx}
\usepackage{hyperref}
\usepackage{color}
\usepackage{xcolor}
\fi
\begin{document}
\let\emph\textit

\title{
Spin Seebeck Effect in Nonmagnetic Excitonic Insulators
  }
\author{Joji Nasu}
\affiliation{
  Department of Physics, Yokohama National University, Hodogaya, Yokohama 240-8501, Japan
}
\author{Makoto Naka}
\affiliation{
  Waseda Institute for Advanced Study, Waseda University, Tokyo 169-8050, Japan
}

\date{\today}
\begin{abstract}
  We propose a mechanism of the spin Seebeck effect attributed to excitonic condensation in a nonmagnetic insulator.
  We analyze a half-filled two-orbital Hubbard model with a crystalline field splitting in the strong coupling limit.
  In this model, the competition between the crystalline field and electron correlations brings about an excitonic insulating state, where the two orbitals are spontaneously hybridized.
  Using the generalized spin-wave theory and Boltzmann transport equation, we find that a spin current generated by a thermal gradient is observed in the excitonic insulating state without magnetic fields.
  The spin Seebeck effect originates from spin-split collective excitation modes although the ground state does not exhibit any magnetic orderings.
  This peculiar phenomenon is inherent in the excitonic insulating state, whose order parameter is time-reversal odd and yields a spin splitting for the collective excitation modes. 
  We also find that the spin current is strongly enhanced and its direction is inverted in the vicinity of the phase transition to another magnetically ordered phase.
  We suggest that the present phenomenon is possibly observed in perovskite cobaltites with the GdFeO$_3$-type lattice distortion.
\end{abstract}
\maketitle


Spin current generation in insulating magnets has attracted considerable attention not only in modern condensed matter physics but also for applications to spintronic devices.
While electric current cannot be produced in insulating magnets, spin current is successfully created by an applied thermal gradient, which is known as the spin Seebeck effect (SSE)~\cite{Xiao2010,Adachi2011,Rezende2014}.
The spin current is carried by spin-polarized collective excitations from a spin aligned ground state~\cite{Maekawa2013rev,adachi2013theory}.
Therefore, materials showing the SSE has been explored in ferrimagnets and ferromagnets~\cite{uchida2010spin,uchida2010,Zhang_magnon2012,Zhang_magnon_full2012}.
Recently, antiferromagnetic insulators and quantum spin liquid under magnetic fields are also considered as the candidates~\cite{qiu2018spin,naka2019spin,hirobe2017one,Minakawa2020}.
In contrast to these magnets, the SSE in nonmagnetic insulators remains elusive because spin-polarized excitations are not trivially present.
Here, we focus on excitonic condensation instead of magnetic orderings to propose another type of the SSE.

The excitonic insulating (EI) state, where the conduction and valence bands are spontaneously hybridized by electron correlations, is a long-standing subject in condensed matter physics~\cite{mott1961transition,PhysRev.158.462,RevModPhys.40.755,PhysRevB.62.2346,PhysRevLett.89.166403,0953-8984-27-33-333201,doi:10.1143/JPSJ.31.730,doi:10.1143/JPSJ.31.812,doi:10.1143/JPSJ.44.1759}.
This state has been proposed to be realized in several multi-orbital materials, e.g., transition metal oxides and chalcogenides~\cite{PhysRevLett.103.026402,Wakisaka2012,PhysRevB.87.035121,PhysRevB.90.245144,PhysRevB.90.235112,PhysRevB.89.115134,Sotnikov2017,Ikeda2016,ikeda2020}.
Recently, it has been theoretically suggested that a carrier-doped excitonic magnet shows spin-split Fermi surfaces, which leads to spin current generation mediated by spin-polarized electrons~\cite{Kunes2016Spontaneous,Geffroy2018,Nishida2019,Yamamoto2020}.
This does not originate from not the intrinsic spin-orbit coupling but is caused by the double-exchange mechanism~\cite{Kunes2016Spontaneous}.
On the other hand, this mechanism is absent in insulators.
Meanwhile, the EI state is characterized by a certain order, which is not accompanied by magnetic orderings but is time-reversal broken~\cite{PhysRevB.90.235112,PhysRevB.89.115134,Nasu2016EI,Kaneko2016multipole,Altarawneh2012,Tatsuno2016,sotnikov2016field,Sotnikov2017}.
Its order parameter gives rise to an effective internal field, which is expected to yield a spin splitting in the excitation spectra.
However, spin transport properties in an undoped EI is not fully elucidated.

In this Letter, we investigate the spin excitations and transport properties of an EI state.
We analyze an effective model derived from the two-orbital Hubbard model in the strong correlation limit, by using the generalized spin-wave theory (GSWT) and Boltzmann transport equation.
We find that the collective spin excitations show a spin-dependent splitting in the nonmagnetic EI (NEI) state.
As the result, a spin conductivity with respect to a thermal gradient becomes nonzero, i.e., the SSE appears, even without magnetic fields.
This is attributed to the time-reversal symmetry breaking inherent in the NEI state.
The spin conductivity substantially increases and its sign changes by temperature near the phase boundary to another EI state because of the softening of the spin-split mode.
Finally, we propose how to verify our mechanism by presenting perovskite cobaltites with the GdFeO$_3$-type lattice distortion.


We start from the following two-orbital Hubbard model~\cite{PhysRevLett.99.126405}: ${\cal H}={\cal H}_U+{\cal H}_t$, where the Hamiltonians for local contributions and intersite electron hoppings are given by
${\cal H}_U=\Delta\sum_i n_{i a} +U\sum_{i \gamma}n_{i\gamma\uparrow}n_{i\gamma\downarrow}+U'\sum_{i} n_{ia}n_{ib}
+J\sum_{i\sigma\sigma'}c_{ia\sigma}^\dagger c_{ib\sigma'}^\dagger c_{ia\sigma'}c_{ib\sigma}
 +I\sum_{i,\gamma\neq \gamma'}c_{i\gamma\uparrow}^\dagger c_{i\gamma \downarrow}^\dagger c_{i\gamma'\downarrow}c_{i\gamma'\uparrow}$
and
 ${\cal H}_t=\sum_{\means{ij}\gamma\sigma}t_\gamma (c_{i\gamma \sigma}^\dagger c_{j\gamma \sigma}+{\it H.c.})
 +V\sum_{\means{ij}\sigma} (c_{ia \sigma}^\dagger c_{jb \sigma}+ c_{ib \sigma}^\dagger c_{ja \sigma}+{\it H.c.})$,
respectively.
Here, $c_{i\gamma \sigma}^\dagger$ is the creation operator of the electron with spin $\sigma(=\uparrow, \downarrow)$ in orbital $\gamma(=a,b)$ at site $i$, and $n_{i\gamma}=\sum_\sigma c_{i\gamma \sigma}^\dagger c_{i\gamma \sigma}$ is the number operator. 
The crystalline field splitting, intraorbital and interorbital Coulomb interactions, Hund coupling, and pair hopping interaction are represented by $\Delta$, $U$, $U'$, $J$, and $I$, respectively.
In addition to the transfer integral $t_\gamma$ between the $\gamma$ orbitals in the nearest neighbor (NN) sites $\means{ij}$ in ${\cal H}_t$, we consider the interorbital hopping $V$ between the different orbitals in the NN sites.

In the present study, we focus on the electronic properties of the half-filling case in the strong correlation limit.
The low-energy local eigenstates for ${\cal H}_U$ are the low-spin (LS) state $\kets{L}=\left(f c_{b\uparrow}^\dagger c_{b\downarrow}^\dagger - g c_{a\uparrow}^\dagger c_{a\downarrow}^\dagger\right)\kets{\emptyset}$, where $f=\sqrt{(1+\Delta/\Delta')/2}$ and $g=\sqrt{(1-\Delta/\Delta')/2}$ with $\Delta'=\sqrt{\Delta^2-I^2}$, and three high-spin (HS) states $\{\kets{X},\kets{Y},\kets{Z}\}$, which are given by
$\kets{X}=\frac{1}{\sqrt{2}}\left(-c_{a\uparrow}^\dagger c_{b\uparrow}^\dagger + c_{a\downarrow}^\dagger c_{b\downarrow}^\dagger\right)\kets{\emptyset}$, $\kets{Y}=\frac{i}{\sqrt{2}}\left(c_{a\uparrow}^\dagger c_{b\uparrow}^\dagger + c_{a\downarrow}^\dagger c_{b\downarrow}^\dagger\right)\kets{\emptyset}$, and 
$\kets{Z}=\frac{1}{\sqrt{2}}\left(c_{a\uparrow}^\dagger c_{b\downarrow}^\dagger + c_{a\downarrow}^\dagger c_{b\uparrow}^\dagger\right)\kets{\emptyset}$.
Here, $\kets{\emptyset}$ stands for the vacuum. 
The effective model is defined for the subspace composed of the direct product of these four local states.
Using the second-order perturbation expansion with respect to ${\cal H}_t$, the low-energy Hamiltonian is obtained as
\begin{align}
  {\cal H}_{\rm eff}&=-\Delta_z\sum_i\tau_i^z+J_{z}\sum_{\means{ij}}\tau_i^z\tau_j^z
  +J_s\sum_{\means{ij}}\bm{S}_i\cdot\bm{S}_j
  \nonumber\\
  &
  -J_x\sum_{\means{ij}\Gamma}\tau_{\Gamma i}^x\tau_{\Gamma j}^x
  -J_y\sum_{\means{ij}\Gamma}\tau_{\Gamma i}^y\tau_{\Gamma j}^y
  -K\sum_{\means{ij}\Gamma} \left(S_{i}^\Gamma \tau_{\Gamma j}^x +\tau_{\Gamma i}^x S_{j}^\Gamma\right),
  \label{eq:hamil-K}
\end{align}
where the exchange constants, $\Delta_z$, $J_s$, $J_x$, $J_y$, and $K$ are determined by the parameters in ${\cal H}$~\cite{suppl}.
In the effective model, $S^X_i$, $S^Y_i$, and $S^Z_i$ represent the spin-1 operators at site $i$ for the HS states, and the pseudospins $\tau_\Gamma^x$, $\tau_\Gamma^y$ ($\Gamma=X,Y,Z$), and $\tau^z$ describe the matrix elements between LS and HS states, where 
$\tau_\Gamma^x=\kets{L}\bras{\Gamma} + \kets{\Gamma}\bras{L}$,
$\tau_\Gamma^y=i\left(\kets{L}\bras{\Gamma}-\kets{\Gamma}\bras{L}\right)$, and
$\tau^z=\sum_\Gamma\left(\kets{\Gamma}\bras{\Gamma} -\kets{L}\bras{L}\right)$, respectively.
Note that $\tau^z$ represents the energy difference between LS and HS states, and $\tau_\Gamma^x$ and $\tau_\Gamma^y$ yield the hybridization between these states.
Therefore, nonzero expectation values of $\tau_\Gamma^x$ and $\tau_\Gamma^y$ indicate the emergence of the EI state.
In particular, the former (latter) is the time-reversal odd (even) operator, and therefore, $\tau_\Gamma^x$ can couple with the spin operators as shown in Eq.~\eqref{eq:hamil-K}, where the coupling constant $K$ is proportional to $(t_a + t_b)V$; $K$ is nonzero in the presence of the interorbital hopping~\cite{suppl}.
In the following calculations, we study the system with the direct gap, $t_a t_b <0$.
In this case, the exchange constants $J_x$ and $J_y$ satisfy the relation $J_x\gtrsim J_y$ and $J_x$ is positive.
This leads to the ferro-type pseudospin correlation while the spin exchange constant $J_s$ is always antiferromagnetic.

To analyze the spin excitations and transport properties in the Hamiltonian Eq.~\eqref{eq:hamil-K}, we apply the GSWT~\cite{Onufrieva1985,Papanicolaou1988367,doi:10.1143/JPSJ.70.3076,doi:10.1143/JPSJ.72.1216,PhysRevB.60.6584,PhysRevB.88.224404,PhysRevB.88.205110,suppl}.
In this method, the mean-field (MF) approximation is applied, and the Hamiltonian is divided into ${\cal H}_{\rm eff}= \sum_i {\cal H}_i^{\rm MF}+{\cal H}'$; ${\cal H}_i^{\rm MF}$ is the local MF Hamiltonian obtained by the decoupling of the exchange interactions and ${\cal H}'$ is the contribution beyond the MF Hamiltonian.
${\cal H}'$ is given by the interactions between the fluctuation around the MFs, $\delta{\cal O}_i={\cal O}_i-\bras{0;C_i}{\cal O}\kets{0;C_i}$, where $\kets{0;C_i}$ is the local MF ground state of ${\cal H}_i^{\rm MF}$ on sublattice $C_i$ to which site $i$ belongs.
In the GSWT, this is approximated as $\delta{\cal O}_i\simeq \sum_n \bras{n;C_i}{\cal O}\kets{0;C_i} a_{in}^\dagger + {\it H.c.}$, where $a_{in}^\dagger$ is a creation operator of a boson, where the summation for $n$ is taken for the local excited states of ${\cal H}_i^{\rm MF}$.
By the above procedure, ${\cal H}_{\rm eff}$ is approximated as ${\cal H}_{\rm SW}$, which is written as a bilinear form of the bosons $a_{in}^\dagger$~\cite{suppl}.
This is diagonalized by the Bogoliubov transformation~\cite{COLPA1978327}.
We introduce a new bosonic operator $\alpha_{\bm{q}\eta}^\dagger$ with the excitation energy $\omega_{\bm{q}\eta}$ for the wave vector $\bm{q}$ and blanch $\eta$. 
Using the GSWT, we calculate the dynamical spin correlator~\cite{Nasu2016EI,suppl}
\begin{align}
  {\cal S}^{\Gamma\Gamma'}(\bm{q},\omega)
  &=\frac{1}{2\pi}\int_{-\infty}^\infty dt \brasd{0}\delta S_{\bm{q}}^{\Gamma}(t) \delta S_{-\bm{q}}^{\Gamma'}\ketsd{0}e^{i\omega t-\delta |t|},
\end{align}
where $\delta S_{\bm{q}}^{\Gamma}=N^{-1/2}\sum_{i}\delta S_{i}^{\Gamma} e^{-i\bm{q}\cdot \bm{r}_i}$, 
${\cal O}(t)=e^{i{\cal H}_{\rm SW}t}{\cal O}e^{-i{\cal H}_{\rm SW}t}$, $\delta$ is a broadening factor, and $\ketsd{0}$ is the vacuum for the Bogoliubov bosons. 

The thermal conductivity $\kappa\equiv\kappa_E^{xx}$ and spin conductivity with respect to a thermal gradient, $\kappa_s\equiv\kappa_{S^z}^{xx}$, are defined by
$\means{J^\mu_{\cal O}}_{\nabla T}/V=\kappa_{\cal O}^{\mu\nu}\left(-\nabla_\nu T\right)$, where $V$ is the volume, $\means{\cdots}_{\nabla T}$ represents the expectation value in the presence of the thermal gradient, and $\mu,\nu=x,y,z$ stand for the coordinate axes.
The energy current $\bm{J}_E$ is defined from the energy polarization $\bm{P}_E= \sum_i \bm{r}_i h_i$ as $\bm{J}_E=i[{\cal H}_{\rm SW},\bm{P}_E]$, where $h_i$ is composed of the terms involving site $i$ in ${\cal H}_{\rm SW}$~\cite{PhysRevLett.104.066403,Matsumoto2011,Matsumoto2014}. 
The spin current $J^\mu_{S^Z}$ is also defined in a similar manner.
The spin polarization is introduced as $\bm{P}_{S^Z}=\sum_i \bm{r}_i S_i^Z\simeq \sum_{in} \bm{r}_i \bras{n;C_i}\delta S_i^Z\kets{n;C_i}a_{in}^\dagger a_{in}+{\it const}.$ when $S_i^z$ commutes with ${\cal H}_i^{\rm MF}$ because of the absence of the spin-orbit coupling. 
The conductivities are calculated using the Boltzmann equation with the relaxation time approximation~\cite{Rezende2014,Rezende2016,Takashima2018,suppl}.
We have numerically confirmed that the results are consistent with those obtained by the Kubo formula~\cite{Ogata2017,naka2019spin}.

\begin{figure}[t]
  \begin{center}
    \includegraphics[width=\columnwidth,clip]{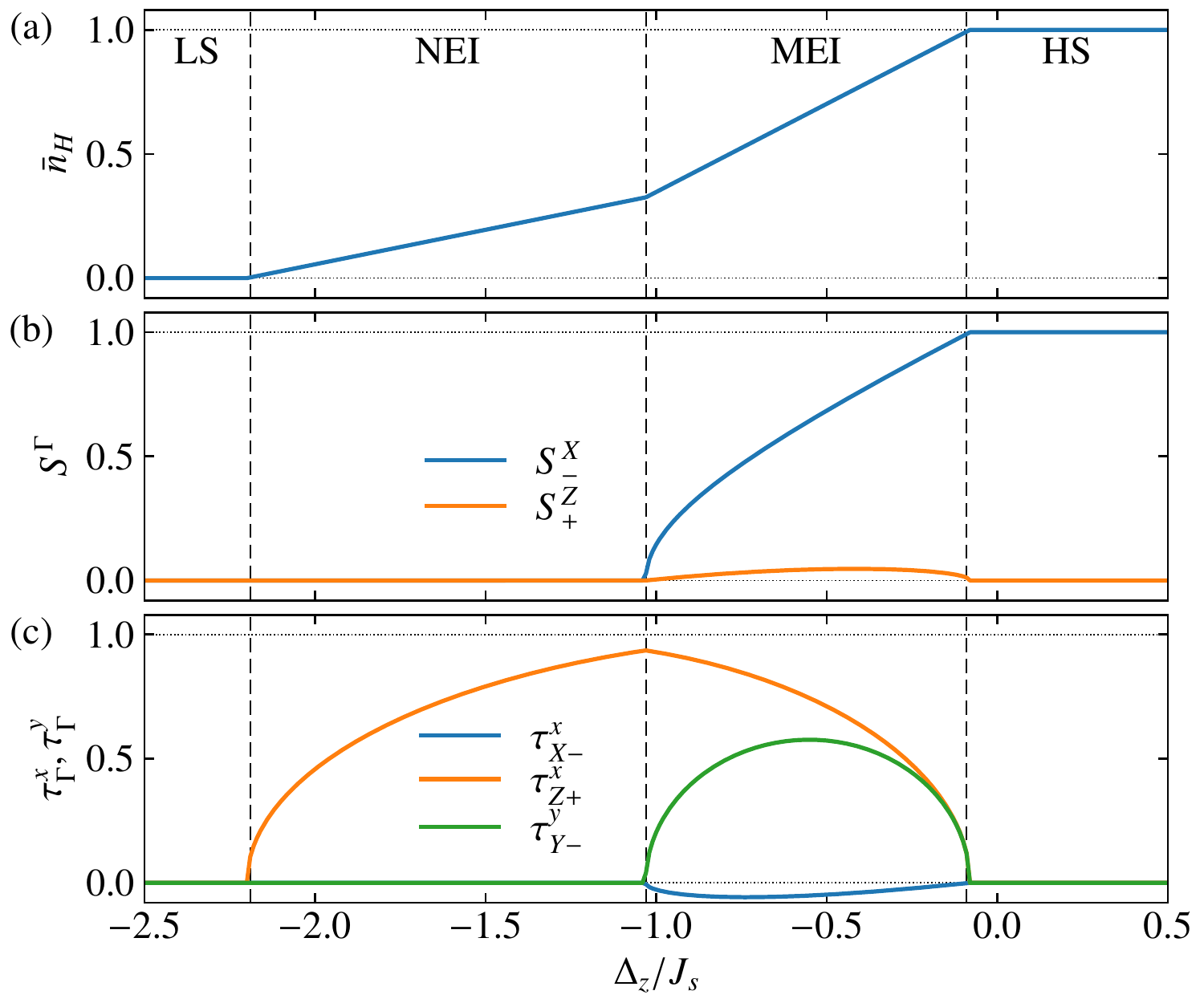}
    \caption{
    (a) High-spin density $\bar{n}_H$, (b) spin moments $S_{\pm}^\Gamma$, and pseudospin moments, $\tau_{\Gamma\pm}^x$ and $\tau_{\Gamma\pm}^y$, as functions of $\Delta_z$ with the exchange parameters $(J_x,J_y,J_z,K)/J_s=(0.5,0,0.1,0.1)$. The dashed lines indicate the phase boundaries.
  }
    \label{fig_mf}
  \end{center}
\end{figure}

\begin{figure*}[t]
  \begin{center}
    \includegraphics[width=2\columnwidth,clip]{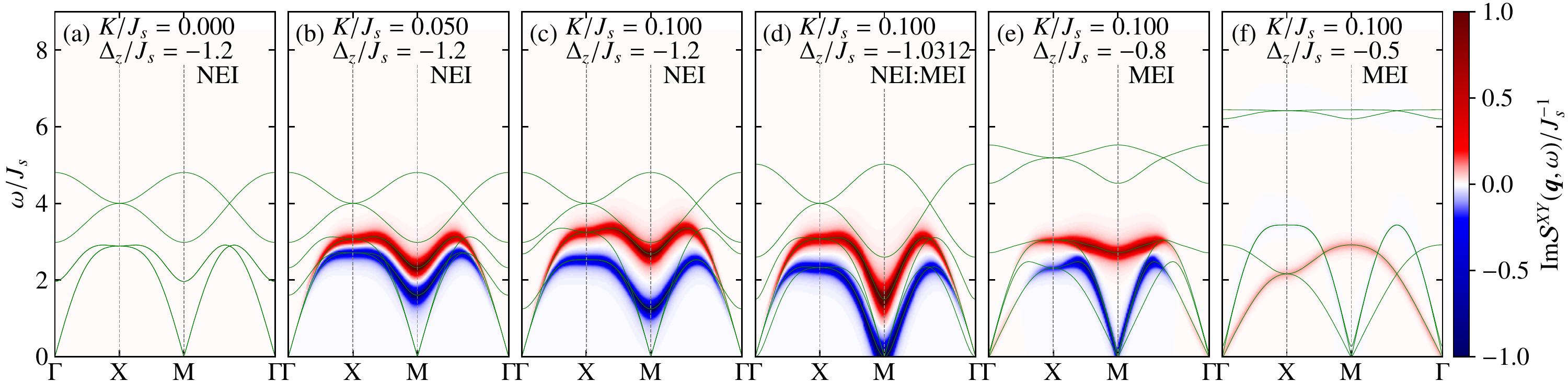}
    \caption{
    Contour plots of the imaginary parts of the dynamical spin correlator, ${\rm Im}{\cal S}^{XY}(\omega)$, with $\delta=0.1J$ at several parameters.
    Green lines represent the excitation energies.
  }
    \label{fig_spec}
  \end{center}
\end{figure*}

First, we show the results of the two-sublattice MF approximation for Eq.~\eqref{eq:hamil-K} on a square lattice, where the length of the primitive translation vectors is set to be unity.
The ground-state MF phase diagrams without the interorbital hopping, i.e., at $K=0$, have been already examined in Ref.~\cite{Nasu2016EI}.
In the present calculations, we choose the exchange parameters as $(J_s, J_x,J_y,J_z)/J_s=(1,0.5,0,0.1)$.
Figure~\ref{fig_mf} shows the $\Delta_z$ dependences of the HS density $\bar{n}_H$ and the spin and pseudospin moments at $K/J_s=0.1$.
We find the four phases, the uniform LS, HS with the AFM order, and two-types of EI phases: the NEI and magnetic EI (MEI)~\footnote{The NEI and MEI were introduced in Ref.~\cite{Nasu2016EI} as EIQ and EIM, respectively.}.
Here, the uniform and staggard spin moments are introduced as $S_{\pm}^\Gamma=\frac{1}{2}\left(\means{S^\Gamma}_A\pm\means{S^\Gamma}_B\right)$ with $\means{S^\Gamma}_A$ and $\means{S^\Gamma}_B$ being the moments on the sublattices $A$ and $B$, respectively.
The pseudospin moments, $\tau_{\Gamma\pm}^x$ and $\tau_{\Gamma\pm}^y$, are defined in the same manner.
The LS (HS) phase is characterized by $\bar{n}_H=0$ ($\bar{n}_H=1$) as shown in Fig.~\ref{fig_mf}(a) and $\bar{n}_H$ continuously changes in the NEI and MEI phases.
In the HS phase, $S_{-}^X=1$, indicating the AFM order for $S^X$ [Fig.~\ref{fig_mf}(b)].
The MEI phase also possesses nonzero $S_{-}^X$ and small FM $S^Z$ components.
Accompanied by the spin canting, $\tau_{X\pm}^x$ takes a small value [Fig.~\ref{fig_mf}(c)] in the MEI state while it is zero at $K=0$.
On the other hand, in the NEI phase, $\tau_{Z\pm}^x$ is only finite, similar to the case with $K=0$.
We find the phase boundaries are almost unchanged by the introduction of $K$.

While the MF ground state in the NEI is not changed qualitatively by $K$, we find the substantial change of the spin excitation spectrum.
As shown in Fig.~\ref{fig_spec}(a), there are four excitation modes in the NEI phase at $K=0$.
The low-energy two gapless modes and high-energy two gapped modes correspond to the spin and orbital excitations, respectively~\cite{Nasu2016EI}.
To examine the spin dependence of the collective modes, we calculate the imaginary part of the dynamical spin correlator ${\cal S}^{XY}(\bm{q},\omega)$~\cite{Barker2016}.
Note that ${\rm Im} {\cal S}^{XY}(\bm{q},\omega) =[{\cal S}^{+-}(\bm{q},\omega)-{\cal S}^{-+}(\bm{q},\omega)]/4$, where $S^{\pm}=S^X\pm i S^Y$.
This expression clearly indicates that the positive (negative) spectral weight corresponds to the spin excitation associated with a negative (positive) change of $S^Z$.
Figure~\ref{fig_spec} shows the contour map of ${\rm Im}{\cal S}^{XY}(\omega)$ and dispersion relations of the collective excitations for the several values of $K$ and $\Delta_z$. 
As shown in Figs.~\ref{fig_spec}(a)--\ref{fig_spec}(c), in the NEI phase with $\Delta_z/J_s=-1.2$, one of the spin excitation modes splits into two by the introduction of $K$.
These two modes are associated with the positive and negative weights of ${\rm Im}{\cal S}^{XY}(\bm{q},\omega)$.
This indicates that the spin splitting of the collective modes is caused by the interorbital hopping, i.e., $K$, although the EI state remains nonmagnetic [see Fig.~\ref{fig_mf}(b)].
In the NEI state, the uniform pseudospin moment for $\means{\tau^x_Z}$ is nonzero, resulting in the effective magnetic field for $S^Z$ by the last term of Eq.~\eqref{eq:hamil-K}.
This effective field does not induce any local spin moments in the ground state but gives rise to the spin splitting in the excited states.

We also find that the spin-split collective modes are softened while $\Delta_z$ approaching the critical point between the NEI and MEI phases, $\Delta_z^{\rm critical}\simeq \Delta_z=1.0312J_s$ [Figs.~\ref{fig_spec}(c) and \ref{fig_spec}(d)].
At this point, one of these modes is gapless at the M point in the Brillouin zone with a quadratic dispersion.
In the MEI phase above $\Delta_z^{\rm critical}$, and two gapless linear dispersions with different velocities appear~\cite{Nasu2016EI}.
In the MEI phase, while increasing $\Delta_z$, the low-energy weight of ${\rm Im}{\cal S}^{XY}(\omega)$ turns from negative to positive [Figs.~\ref{fig_spec}(e) and \ref{fig_spec}(f)].

\begin{figure}[t]
  \begin{center}
    \includegraphics[width=\columnwidth,clip]{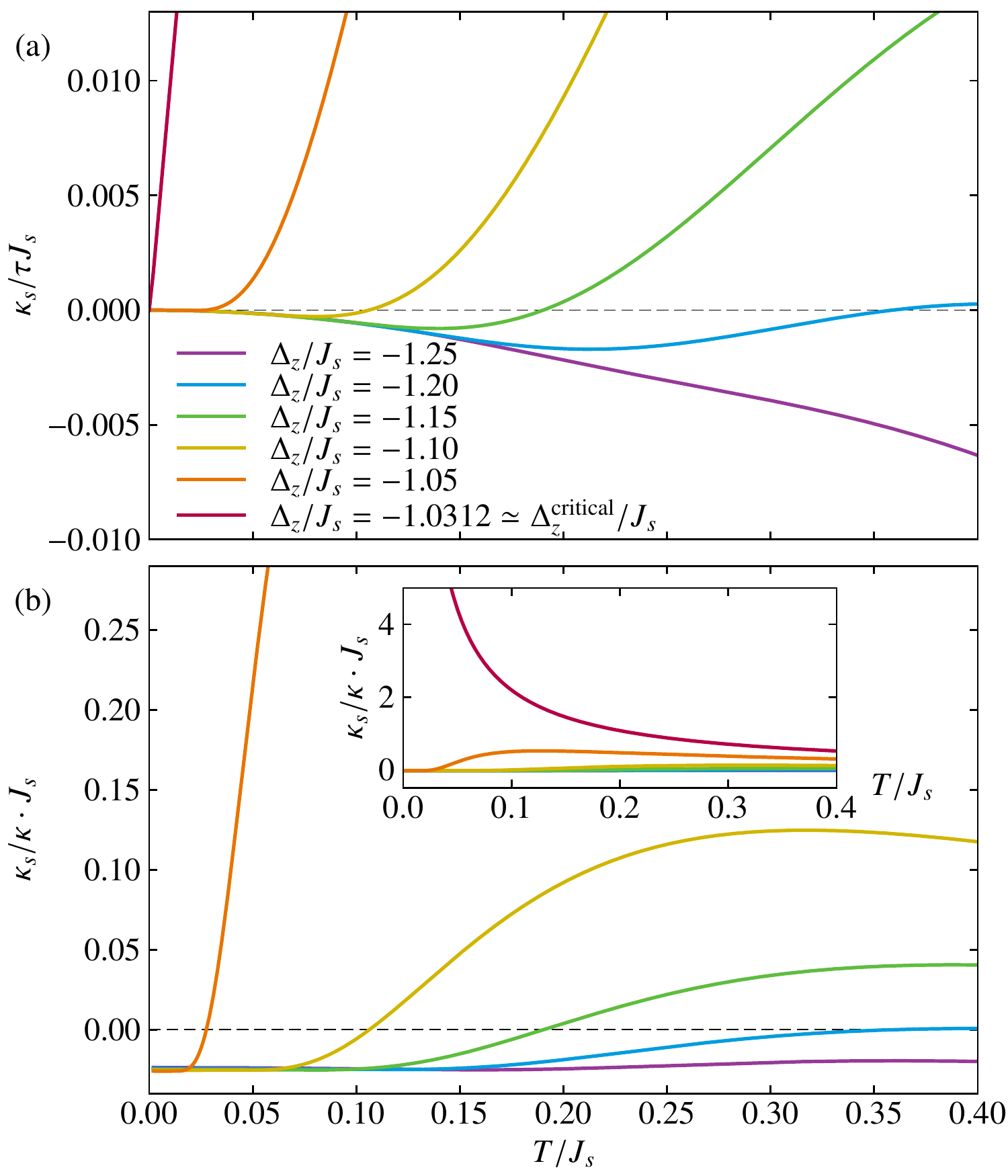}
    \caption{
    Temperature dependences of (a) the spin conductivity $\kappa_s$ and (b) the ratio $\kappa_s/\kappa$ at several $\Delta_z$ in the NEI phase with the exchange parameters $(J_x,J_y,J_z,K)/J_s=(0.5,0,0.1,0.1)$, where $\tau$ is the relaxation time.
    The inset of (b) is its extended plot.
  }
    \label{fig_current}
  \end{center}
\end{figure}

Keeping this in mind, let us examine spin transport properties in the EI phases when a thermal gradient is applied to the system.
Figure~\ref{fig_current}(a) shows the temperature dependence of the spin conductivity $\kappa_s$ in the NEI phase~\cite{footnote_excitonic-sc}.
In the NEI phase at $\Delta_z=-1.25$ far from $\Delta_z^{\rm critical}$, $\kappa_s/\tau$ is negative and decreases with increasing temperature.
This is understood as follows.
As shown in Figs.~\ref{fig_spec}(b) and \ref{fig_spec}(c), the group velocity of the collective mode with the positive weight of ${\rm Im}{\cal S}^{XY}(\omega)$ is larger than that with the negative weight around the $\Gamma$ point.
This implies that the spin excitation decreasing $S^Z$ contributes dominantly to the spin transport, and hence, $\kappa_s$ is negative.

At $\Delta_z=-1.2$, $\kappa_s/\tau$ turns to increase with increasing temperature and its sign changes by the temperature evolution.
Further increase of $\Delta_z$ enhances the spin conductivity strongly in the high temperature region.
At the critical point $\Delta_z^{\rm critical}$, $\kappa_s/\tau$ largely increases proportional to temperature as shown in the inset of Fig.~\ref{fig_current}(a).
The peculiar temperature dependence is attributed to the softening of the spin-split collective mode at the M point as shown in Fig.~\ref{fig_spec}(d).
This mode is associated with the negative ${\rm Im}S^{XY}(\omega)$, indicating the excitation with raising $S^Z$, and therefore, $\kappa_s$ becomes positive near $\Delta_z^{\rm critical}$.
In the MEI phase, our formalism is not applicable for calculating $\kappa_s$~\cite{footnote_excitonic-sc} but we expect that $\kappa_s$ changes to positive to negative while increasing $\Delta_z$ from $\Delta_z^{\rm critical}$ on the basis of the low energy behavior of ${\rm Im}{\cal S}^{XY}(\omega)$ shown in Figs.~\ref{fig_spec}(d)--\ref{fig_spec}(f).

To examine the conversion ratio from the thermal to spin current, we calculate $\kappa_s/\kappa$ as shown in Fig.~\ref{fig_current}(b).
At $\Delta_z/J_s=-1.25$, this quantity is negative but the sign change is seen in the case close to $\Delta_z^{\rm critical}$.
While $\kappa_s/\kappa$ approaches a negative common value in the low temperature limit in the NEI phase, the different behavior is seen at $\Delta_z^{\rm critical}$; $\kappa_s/\kappa$ appears to diverge with decreasing temperature as shown in the inset of Fig.~\ref{fig_current}(b).

\begin{figure}[t]
  \begin{center}
    \includegraphics[width=\columnwidth,clip]{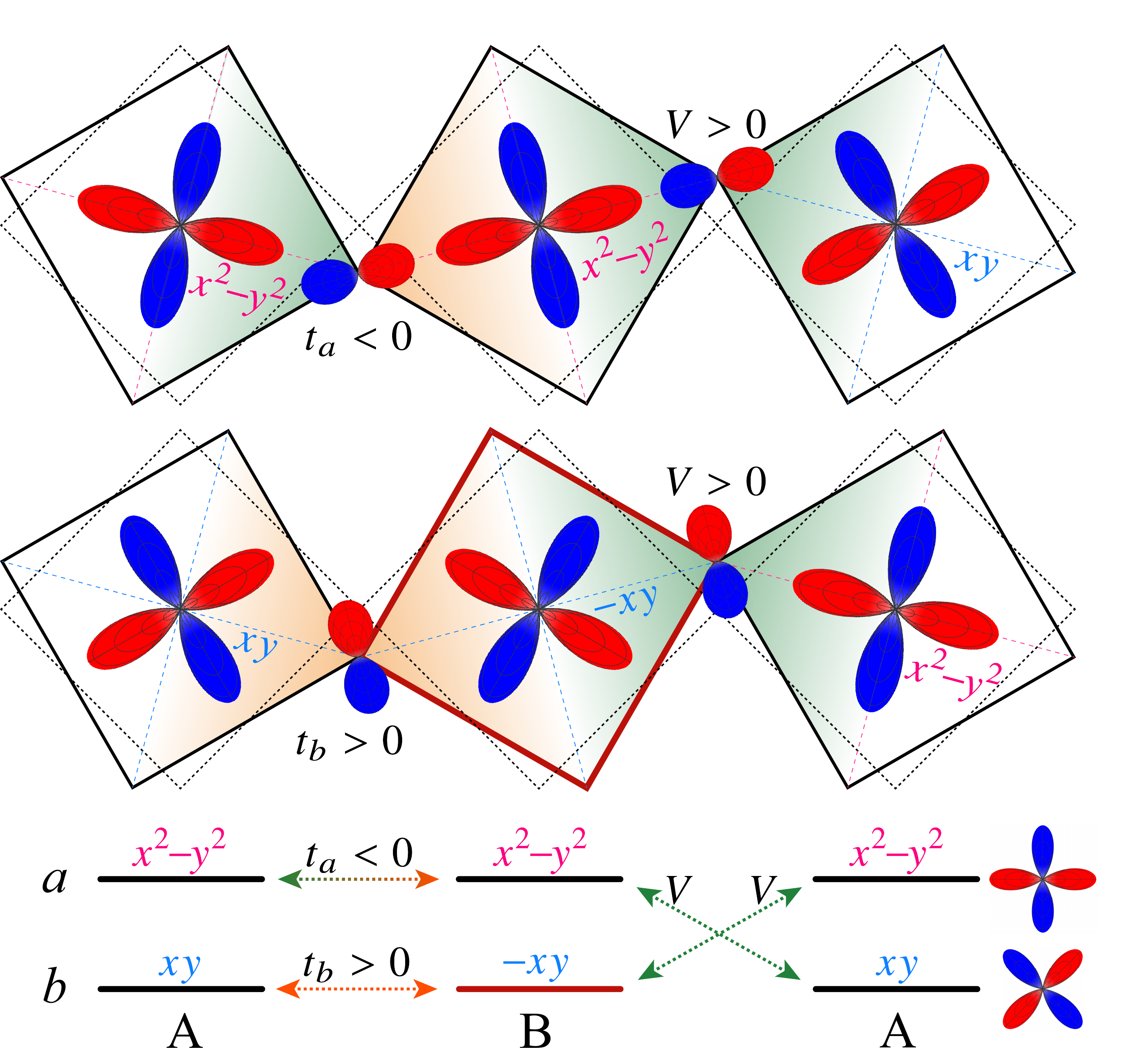}
    \caption{
Schematic pictures of orbital configurations in the GdFeO$_3$-type distortion to understand the correspondence to the present model.
The $p$ orbitals are also depicted on the coordinate of the middle octahedra.
The gauge transformation for the $d_{xy}$ orbital to $-d_{xy}$ are applied to the octahedron surrounded by the red line. 
The orange and green colors in the octahedra represent the positive and negative $d$-$p$ hybridizations, respectively.
Transfer integrals between the same and different orbitals are shown in the bottom.
  }
    \label{fig_orbital}
  \end{center}
\end{figure}

Here, we discuss how to verify our theoretical proposal in real materials.
One of the candidates of EIs is the perovskite cobaltite Pr$_{0.5}$Ca$_{0.5}$CoO$_3$~\cite{PhysRevLett.103.026402,Wakisaka2012,PhysRevB.87.035121,PhysRevB.90.245144}.
This material exhibits a metal-insulator transition at about 90~K and magnetic orderings have not been observed experimentally.
The Co ions are expected to be trivalent and multiple spin-states such as LS ($t_{2g}^6$) and HS ($t_{2g}^4 e_{g}^2$) are energetically competing.
It was suggested that the NEI state with the spontaneous hybridization occurs between $d_{x^2-y^2}$ and $d_{xy}$ orbitals by first principles calculations~\cite{PhysRevB.90.235112}.
Therefore, the present two-orbital Hubbard model is a minimal model to capture the nature of the EI state in the material.
The space group of Pr$_{0.5}$Ca$_{0.5}$CoO$_3$ is orthorhombic ($Pnma$) with the GdFeO$_3$-type distortion in both higher and lower temperatures~\cite{PhysRevB.66.052418,Tsubouchi2004,Hejtmanek2010}.
The detailed structure analysis suggested that the rotation of CoO$_6$ octahedra corresponding to the GdFeO$_3$-type distortion is enhanced with decreasing temperature.
This distortion gives rise to the intersite interorbital hopping between the $d_{x^2-y^2}$ in the $e_g$ orbitals and $d_{xy}$ in the $t_{2g}$ orbitals as shown in Fig.~\ref{fig_orbital}.
This effect is incorporated as $V$ in the two-orbital Hubbard model.
By applying the gauge transformation for the $b$ orbital in the $B$ sublattice, the Hamiltonian is mapped onto the case with the uniform interorbital hopping and $t_a t_b<0$, which is nothing but the system addressed by the present study.
This transformation does not affect the spin operators and eigenenergies, and therefore, the results for the spin-split excitations and SSE
are expected to be observed in the real material Pr$_{0.5}$Ca$_{0.5}$CoO$_3$ if the excitonic order is realized.

As shown in Fig.~\ref{fig_current}(b), despite the absence of magnetic orders and magnetic fields, the order of $\kappa_s$ is close to that of $\kappa$ in natural units, which is similar to the case of ferromagnetic Heisenberg models.
Since the SSE mediated by the collective excitations has been measured in ferromagnetic/ferrimagnetic insulators such as iron-based garnets~\cite{uchida2010spin,uchida2010}, we expect that the spin current is observed when the EI state is realized in the candidate material.
Furthermore, in the iron-based garnets, spin-dependent magnons were observed by inelastic neutron scattering measurements~\cite{Nambu2020}.
Therefore, we believe that the SSE and spin splitting in collective excitations can be measured experimentally in candidate materials of the EI state although its order parameter is a higher-order multipole~\cite{PhysRevB.90.235112,Nasu2016EI,Kaneko2016multipole}, which is difficult to be observed by conventional probes.

In summary, we have proposed a mechanism of the spin Seebeck effect in a nonmagnetic excitonic insulating state by analyzing the effective model derived from the two-orbital Hubbard model in strong correlation limit.
We have revealed that the spin Seebeck effect originates from the spin-split collective excitations, which is caused by an internal effective field emerging from the excitonic order parameter.
We have also suggested that these phenomena will be experimentally observed in perovskite cobaltites with the GdFeO$_3$-type distortion.

\begin{acknowledgments}
The authors thank S.~Ishihara, Y.~Ohta, K.~Sugimoto, and S.~Yamamoto for fruitful discussions.
Parts of the numerical calculations were performed in the supercomputing
systems in ISSP, the University of Tokyo.
  This work was supported by Grant-in-Aid for Scientific Research from
  JSPS, KAKENHI Grant No.~JP19K03723.
\end{acknowledgments}

\bibliography{refs}

\begin{thebibliography}{63}
\bibitem [{foo()}]{footnote_excitonic-sc}%
  \BibitemOpen
  \href@noop {} {}\bibinfo {note} {Note that $\kappa_s$ is valid only for the
  case with the conservation of the total $S^Z$ in ${\cal H}_{\rm
  SW}$~\cite{suppl}. We calcualte this in the NEI phase but do not in the MEI
  phase because the spin current does not a conserved current in the MEI phase
  due to the small spin canting.}\BibitemShut {Stop}%
\end{thebibliography}

\clearpage
\onecolumngrid

\appendix
\vspace{15pt}
\begin{center}
{\large \bf Supplemental Material for ``Spin Seebeck Effect in Nonmagnetic Excitonic Insulators''}
\end{center}

\setcounter{figure}{0}
\setcounter{equation}{0}
\setcounter{table}{0}
\renewcommand{\thefigure}{S\arabic{figure}}
\renewcommand{\theequation}{S\arabic{equation}}
\renewcommand{\thetable}{S\Roman{table}}
\baselineskip=6mm

\section{Deviation of effective Hamiltonian}

In this section, we derive the low-energy effective Hamiltonian from the two-orbital Hubbard model given by
\begin{align}
 {\cal H}={\cal H}_U+{\cal H}_t.\label{eq:sup_hamil}
\end{align}
Here, the local part is represented as
\begin{align}
 {\cal H}_U=\Delta\sum_i n_{i a} +U\sum_{i \gamma}n_{i\gamma\uparrow}n_{i\gamma\downarrow}+U'\sum_{i} n_{ia}n_{ib}
 +J\sum_{i\sigma\sigma'}c_{ia\sigma}^\dagger c_{ib\sigma'}^\dagger c_{ia\sigma'}c_{ib\sigma}
 +I\sum_{i\gamma\neq \gamma'}c_{i\gamma\uparrow}^\dagger c_{i\gamma \downarrow}^\dagger c_{i\gamma'\downarrow}c_{i\gamma'\uparrow},\label{eq:sup_hamilU}
\end{align}
where $c_{i\gamma \sigma}^\dagger$ is the creation operator of the electron with spin $\sigma(=\uparrow, \downarrow)$ in orbital $\gamma(=a,b)$ at site $i$, and $n_{i\gamma}=\sum_\sigma c_{i\gamma \sigma}^\dagger c_{i\gamma \sigma}$ is the number operator. 
The terms with the coefficients, $\Delta$, $U$, $U'$, $J$, and $I$, represent the crystalline field splitting, intraorbital and interorbital Coulomb interactions, Hund coupling, and pair hopping interaction, respectively.
These coefficients are positive.
The second term of Eq.~\eqref{eq:sup_hamil} represents intersite electron hopping, which is given by
\begin{align}
 {\cal H}_t=\sum_{\means{ij}\gamma\sigma}t_\gamma (c_{i\gamma \sigma}^\dagger c_{j\gamma \sigma}+{\rm H.c.})
 +\sum_{[ij]\sigma} (V_1 c_{ia \sigma}^\dagger c_{jb \sigma}+ V_2 c_{ib \sigma}^\dagger c_{ja \sigma}+{\rm H.c.}),\label{eq:sup_hamilt}
\end{align}
where $t_\gamma$ is the transfer integral between the $\gamma$ orbitals in the nearest neighbor (NN) sites $\means{ij}$, and $V_1$ and $V_2$ are the interorbital hoppings between the different orbitals in the NN sites.
In the latter, we need to beware of the order of sites on each bond, and we introduce $[ij]$, which stands for the ordered NN sites.
The ordering manner is defined by Eq.~\eqref{eq:sup_hamilt}.
In the following calculations, the Boltzmann constant $k_B$, reduced Planck constant $\hbar$, and the length of the primitive translation vectors of the lattice on which the model is defined are set to be unity.

The perturbation procedure is based on the strong correlation limit.
Namely, ${\cal H}_U$ and ${\cal H}_t$ are regarded as unperturbed and perturbed terms, respectively.
In the present study, we focus on the system with the half-filling condition, where the averaged electron number is two.
There are six local states where two electrons are present at a site.
Among them, we extract the following four low-energy states from the eigenstates of ${\cal H}_U$.
Three of them are the high-spin (HS) states with the total spin $S=1$, which are given by 
\begin{align}
  \kets{S^Z=+1}&=c_{a\uparrow}^\dagger c_{b\uparrow}^\dagger\kets{\emptyset}\\
  \kets{S^Z=0}&=\frac{1}{\sqrt{2}}\left(c_{a\uparrow}^\dagger c_{b\downarrow}^\dagger + c_{a\downarrow}^\dagger c_{b\uparrow}^\dagger\right)\kets{\emptyset}\\
  \kets{S^Z=-1}&=c_{a\downarrow}^\dagger c_{b\downarrow}^\dagger\kets{\emptyset},
 \end{align}
 where the eigenenergy of ${\cal H}_U$ is $E_H=\Delta+U'-J$, $S^Z$ is the $Z$ component of the total spin, and $\kets{\emptyset}$ is the vacuum.
The other is the low-spin (LS) state with $S=0$, which is given by
\begin{align}
  \kets{L}&=\left(f c_{b\uparrow}^\dagger c_{b\downarrow}^\dagger - g c_{a\uparrow}^\dagger c_{a\downarrow}^\dagger\right)\kets{\emptyset},
 \end{align}
whose eigenenergy is $E_L=U+\Delta-\Delta'$.
Here, $\Delta'=\sqrt{\Delta^2+I^2}$ and the coefficients $f$ and $g$ are given by
\begin{align}
 f&=\left[1+\left(\frac{\Delta-\Delta'}{I}\right)^2\right]^{-1/2}
 =\frac{1}{\sqrt{2}}\sqrt{1+\frac{\Delta}{\Delta'}}
 =\frac{\Delta+\Delta'}{\sqrt{2\Delta'(\Delta+\Delta')}}\\
 g&=\sqrt{1-f^2}
 =\frac{1}{\sqrt{2}}\sqrt{1-\frac{\Delta}{\Delta'}}
 =\frac{J'}{\sqrt{2\Delta'(\Delta+\Delta')}}.
\end{align}
For the HS states, we introduce the following quadrupolar bases:
\begin{align}
  \kets{X}&=\frac{1}{\sqrt{2}}\left(-\kets{S^Z=+1}+\kets{S^Z=-1}\right)\\
  \kets{Y}&=\frac{i}{\sqrt{2}}\left(\kets{S^Z=+1}+\kets{S^Z=-1}\right)\\
  \kets{Z}&=\kets{S^Z=0}.
 \end{align}

By performing the second order perturbation with respect to ${\cal H}_t$, the effective Hamiltonian is calculated using the following expression:
\begin{align}
  \left({\cal H}_{\rm eff}\right)_{\alpha\alpha'}=\bras{\alpha}{\cal H}_U\kets{\alpha'}
  +\frac{1}{2}\sum_\beta\left(
  \frac{\bras{\alpha}{\cal H}_t\kets{\beta}\bras{\beta}{\cal H}_t\kets{\alpha'}}{E_{\alpha}-E_\beta}
  +\frac{\bras{\alpha}{\cal H}_t\kets{\beta}\bras{\beta}{\cal H}_t\kets{\alpha'}}{E_{\alpha'}-E_\beta}
  \right),~\label{eq:supp-gHeff}
\end{align}
where $\kets{\alpha}$ is an eigenstate of ${\cal H}_U$ with the eigenenergy $E_{\alpha}$, which is given as a direct product of local states $\kets{L}$, $\kets{X}$, $\kets{Y}$, $\kets{Z}$.
The intermediate state $\kets{\beta}$ is also an eigenstate of ${\cal H}_U$ with the eigenenergy $E_\beta$.

Using the general expression in Eq.~\eqref{eq:supp-gHeff}, we obtain the low-energy Hamiltonian as follows~\cite{PhysRevB.89.115134,Nasu2016EI}:
\begin{align}
  {\cal H}_{\rm eff}&=\tilde{\Delta}\sum_in^H_i+J_{n}\sum_{\means{ij}}n^H_in^H_j
  +J_s\sum_{\means{ij}}\bm{S}_i\cdot\bm{S}_j\nonumber\\
  &\qquad+\sum_{\means{ij}}\left(J'\bm{d}_i^\dagger \cdot \bm{d}_{j}^\dagger + {\rm H. c.}\right)
  +\sum_{\means{ij}}\left(J''\bm{d}_i^\dagger \cdot\bm{d}_{j} + {\rm H. c.}\right)
  +\sum_{[ij]}\left(K_{\dashv} \bm{S}_{i}\cdot \bm{d}_j^\dagger + K_{\vdash} \bm{d}_i^\dagger \cdot\bm{S}_{j} + {\rm H. c.}\right),
 \end{align}
where the constant terms are omitted.
The operator $n^H_i$ is the number operator for the HS state at site $i$, $\bm{S}_i=(S^X_i, S^Y_i, S^Z_i)$ represent the spin-1 operators for the HS states, and $\bm{d}_i^\dagger=(d_{Xi}^\dagger, d_{Yi}^\dagger, d_{Zi}^\dagger)$ and $\bm{d}_i^\dagger=(d_{Xi}, d_{Yi}, d_{Zi})$ are the creation and annihilation operators of the excitons, where $d_{\Gamma}^\dagger=\kets{\Gamma}\bras{L}$ and $d_{\Gamma}=\kets{L}\bras{\Gamma}$ with $\Gamma=X,Y,Z$.
The coefficients are given by
\begin{align}
  \tilde{\Delta}&=E_H-E_L+z(\delta E_{LL}-\delta E_{LH})\\
  J_n&=2\delta E_{LH}-\delta E_{LL}-J_s\\
  J_s&=\frac{t_a^2+t_b^2}{E_1-2E_H}+\frac{|V_1|^2+|V_2|^2}{2}\left(\frac{1}{E_2-2E_H}+\frac{1}{E_3-2E_H}\right),\label{eq:supp-Jsex}\\
  J'&=2 fg t_a t_b\left[\frac{1}{E_1-2E_H}+\frac{1}{E_1-2E_L}\right]+f^2 V_1 V_2^*\left[\frac{1}{E_2-2E_H}+\frac{1}{E_2-2E_L}\right]+g^2 V_1^* V_2\left[\frac{1}{E_3-2E_H}+\frac{1}{E_3-2E_L}\right],\label{eq:supp-Jpex}\\
  J''&=2t_a t_b\left[\frac{f^2}{E_2-(E_L+E_H)}+\frac{g^2}{E_3-(E_L+E_H)}\right]+\frac{2fg(V_1V_2^*+V_1^*V_2)}{E_1-(E_L+E_H)},\label{eq:supp-Jppex}\\
  K_{\dashv} &=\frac{f (t_a V_1 +t_b V_2^*)+g (t_a V_1^* +t_b V_2)}{2\sqrt{2}}\left[\frac{1}{E_1-2E_H}+\frac{1}{E_1-(E_L+E_H)}\right]\nonumber\\
  &\qquad +\frac{f (t_a V_1 +t_b V_2^*)}{2\sqrt{2}}\left[\frac{1}{E_2-2E_H}+\frac{1}{E_2-(E_L+E_H)}\right] +\frac{g (t_a V_1^* +t_b V_2)}{2\sqrt{2}}\left[\frac{1}{E_3-2E_H}+\frac{1}{E_3-(E_L+E_H)}\right],\label{eq:supp-K1ex}\\
  K_{\vdash} &=\frac{f (t_a V_2^* +t_b V_1)+g (t_a V_2 +t_b V_1^*)}{2\sqrt{2}}\left[\frac{1}{E_1-2E_H}+\frac{1}{E_1-(E_L+E_H)}\right]\nonumber\\
  &\qquad +\frac{f (t_a V_2^* +t_b V_1)}{2\sqrt{2}}\left[\frac{1}{E_2-2E_H}+\frac{1}{E_2-(E_L+E_H)}\right] +\frac{g (t_a V_2 +t_b V_1^*)}{2\sqrt{2}}\left[\frac{1}{E_3-2E_H}+\frac{1}{E_3-(E_L+E_H)}\right],\label{eq:supp-K2ex}
 \end{align}
where
\begin{align}
  \delta E_{LL}&=\frac{4f^2g^2(t_a^2+t_b^2)}{E_1-2E_L}+\frac{2f^4(|V_1|^2+|V_2|^2)}{E_2-2E_L}+\frac{2g^4(|V_1|^2+|V_2|^2)}{E_3-2E_L},\\
  \delta E_{LH}&=\frac{f^2(t_a^2+t_b^2)}{E_2-(E_L+E_H)}+\frac{g^2(t_a^2+t_b^2)}{E_3-(E_L+E_H)}
  +\frac{|V_1|^2+|V_2|^2}{E_1-(E_L+E_H)},
\end{align}
and $z$ is the coordination number of the lattice on which the Hamiltonian is defined.

To simplify the model Hamiltonian, we assume that $V_1$ and $V_2$ are real numbers.
We also introduce the pseudospin operators as
\begin{align}
  \tau_{\Gamma}^x&=d_{\Gamma}+ d_{\Gamma}^\dagger,\\
  \tau_{\Gamma}^y&=i(d_{\Gamma}- d_{\Gamma}^\dagger),\\
  \tau^z&=\sum_{\Gamma}\left(\kets{\Gamma}\bras{\Gamma}-\kets{L}\bras{L}\right)
  =4n^H-3.
\end{align}
Then, the effective Hamiltonian is represented as
\begin{align}
{\cal H}_{\rm eff}=&-\Delta_z\sum_i\tau_i^z+J_{z}\sum_{\means{ij}}\tau_i^z\tau_j^z
+J_s\sum_{\means{ij}}\bm{S}_i\cdot\bm{S}_j
-J_x\sum_{\means{ij}\Gamma}\tau_{\Gamma i}^x\tau_{\Gamma j}^x
-J_y\sum_{\means{ij}\Gamma}\tau_{\Gamma i}^y\tau_{\Gamma j}^y
-\sum_{[ij]\Gamma} \left( K_{sx} S_{i}^\Gamma \tau_{\Gamma j}^x + K_{xs} \tau_{\Gamma i}^x S_{j}^\Gamma\right),
\label{eq:suppl-Hefft}
\end{align}
where the constant terms are omitted and the coefficients are given by
\begin{align}
\Delta_z=-\left(\frac{\tilde{\Delta}}{4}+\frac{3zJ_n}{16}\right),
J_z=\frac{J_n}{16},
J_x=-\frac{J''+J'}{2},
J_y=-\frac{J''-J'}{2},
K_{sx}=-K_{\dashv}, K_{xs}=-K_{\vdash}.
\end{align}

Here, we discuss the sign of the exchange constants in Eq.~\eqref{eq:suppl-Hefft}.
It is clearly shown in Eq.~\eqref{eq:supp-Jsex} that $J_s$ is always positive, indicating the antiferromangetic interaction.
In the case of small interorbital hoppings, $J'$ and $J''$ are approximately proportional to $t_a t_b$ as shown in in Eqs.~\eqref{eq:supp-Jpex} and \eqref{eq:supp-Jppex}.
Moreover, $J'\lesssim J''$ in the case that pair hopping interaction is small.
In the situation, $J_x$ is negative (positive) when $t_a t_b$ is positive (negative), and $|J_x|\gtrsim|J_y|$ is expected.
Moreover, the coupling constants $K_{sx}$, $K_{xs}$ are zero in the absence of the interorbital hopping as shown in in Eqs.~\eqref{eq:supp-K1ex} and \eqref{eq:supp-K2ex}.
When we assume $V_1=V_2=V$, we find that $K_{sx}=K_{xs}$ and these are proportional to $(t_a + t_b)V$.

\section{Generalized spin wave theory}

\subsection{Mean-field theory}

We consider the localized electron model, whose Hamiltonian is generally written as
\begin{align}
 {\cal H}=\sum_{\means{ij}}\sum_{\alpha\beta} J_{ij}^{\alpha\beta} {\cal O}_{\alpha i}{\cal O}_{\beta j}
-\sum_{i}\sum_\alpha h_{\alpha} {\cal O}_{\alpha i},
\end{align}
where $i$ is the site index, and $\alpha$ and $\beta$ are the labels distinguishing the local operators.
The exchange interaction $J_{ij}^{\alpha\beta}$ and field $h_{\alpha}$ are symbolically introduced.
By applying the mean-field (MF) approximation, the MF Hamiltonian is represented as
\begin{align}
  {\cal H}^{\rm MF}=\sum_i {\cal H}_i^{\rm MF} - \frac{zN}{2}\sum_{\means{ij}}\sum_{\alpha\beta} J_{ij}^{\alpha\beta}\means{{\cal O}_{\alpha}}_{C_i}\means{{\cal O}_{\beta}}_{C_j},
\end{align}
where $N$ is the number of sites, and the local Hamiltonian is given by
\begin{align}
  {\cal H}_i^{\rm MF} = \sum_\alpha
  \left( \sum_{j\in {\rm NN:}i}\sum_{\beta}J_{ij}^{\alpha\beta} \means{{\cal O}_{\beta}}_{C_j} -h_{\alpha}\right)
  {\cal O}_{\alpha i}.
\end{align}
Here, $\{{\rm NN}:i\}$ stands for the NN sites of $i$, and $\means{{\cal O}}_{C_i}=\bras{0;C_i}{\cal O}\kets{0;C_i}$ represents the expectation value for the ground state $\kets{0;C_i}$ of the local Hamiltonian ${\cal H}_i^{\rm MF}$, where site $i$ belongs to the sublattice $C_i$.
The contributions beyond the MF approximation, ${\cal H}'={\cal H}-{\cal H}^{\rm MF}$, are represented as
\begin{align}
  {\cal H}'=
 \sum_{\means{ij}}\sum_{\alpha\beta} J_{ij}^{\alpha\beta} \delta{\cal O}_{\alpha i}\delta{\cal O}_{\beta j}.\label{eq:suppl-Hp}
 \end{align}
The deviation from the MF is given by 
\begin{align}
  \delta{\cal O}_{\alpha i}={\cal O}_{\alpha i}-\means{{\cal O}_{\alpha}}_{C_i}.\label{eq:supp-Op}
\end{align}

\subsection{Generalized Holstein-Primakoff transformation}

To take account of the effect of ${\cal H}'$, the local operator $\delta{\cal O}_{\alpha i}$ is expanded on the basis of the eigenstates of $\kets{m;C_i}$ of the local Hamiltonian ${\cal H}_i^{\rm MF}$ on the sublattice $C_i$, where $m=0,1,\cdots {\cal N}$ with ${\cal N}$ being the number of the excited states (${\cal N}+1$ is the number of the local states):
\begin{align}
 \delta{\cal O}_{\alpha i}=\sum_{mm'}\kets{m;C_i}\bras{m;C_i}\delta{\cal O}_{\alpha i}\kets{m';C_i}\bras{m';C_i}=\sum_{mm'}X_i^{mm'}\bras{m;C_i}\delta{\cal O}_{\alpha i}\kets{m';C_i},
\end{align}
where we introduce the projection operator $X_i^{mm'}=\kets{m;C_i}\bras{m;C_i}$.
Note that $\bras{m;C_i}\delta{\cal O}_{\alpha i}\kets{m';C_i}$ depends only on the sublattice index $C$ of $i$.
The projection operator satisfies the following commutation relation:
\begin{align}
  [X_{i}^{mm'}, X_{i'}^{m''m'''}]=\delta_{ii'}(X_{i}^{mm'''}\delta_{m''m'}-X_{i}^{m''m'}\delta_{mm'''}).
 \end{align}
This commutation relation is reproduced by introducing the bosonic operators $a_{in}$ and $a_{in}^\dagger$ ($n=1,2,\cdots {\cal N}$) as follows: 
\begin{align}
  \begin{cases}
    X_{i}^{n0}=a_{in}^{\dagger}\left({\cal M}-\sum_{n'=1}^{\cal N}a_{in'}^{\dagger} a_{in'}\right)^{1/2} ,\quad X_{i}^{0n}=(X_i^{n0})^\dagger & \textrm{for}\ n\geq 1\\
    X_{i}^{nn'}=a_{in}^\dagger a_{in'} & \textrm{for}\ n,n'\geq 1\\
    X_{i}^{00}={\cal M}-\sum_{n=1}^{\cal N}a_{in}^\dagger a_{in},
  \end{cases}\label{eq:supp-HP}
 \end{align}
where ${\cal M}$ is defined by ${\cal M} = X_{i}^{00}+\sum_{n=1}^{\cal N}a_{in}^\dagger a_{in}$ but should be unity because any one of the local states must be occupied by a boson.
This is the generalized Holstein-Primakoff transformation~\cite{Onufrieva1985,Papanicolaou1988367,doi:10.1143/JPSJ.70.3076,doi:10.1143/JPSJ.72.1216,PhysRevB.60.6584}.

\subsection{Generalized spin-wave approximation}

Given the low-temperature and low-energy system, the existence probability of the ground state $\kets{0;C}$ at each site is high and the probability taking excited states is low enough.
In this case, the approximation $\left({\cal M}-\sum_{n'=1}^{\cal N}a_{in'}^{\dagger} a_{in'}\right)^{1/2}\simeq \sqrt{{\cal M}}=1$ is justified in Eq~\eqref{eq:supp-HP}.
Futhermore, we only consider the matrix elements involving the local MF ground state in $\delta{\cal O}_{\alpha i}$ as 
\begin{align}
  \delta{\cal O}_{\alpha i}\simeq \sum_{n=1}^{\cal N} X_i^{n0}\bras{n;C_i}\delta{\cal O}_{\alpha i}\kets{0;C_i} + H.c.\label{eq:supp-Oi}
\end{align}
Hereafter, the summation for $n$ is taken for $n=1,2\cdots {\cal N}$.
Using this approximation, ${\cal H}'$ in Eq~\eqref{eq:suppl-Hp} is represented as
\begin{align}
  {\cal H}'\simeq \sum_{\means{ij}}\sum_{\alpha\beta}\sum_{nn'}J_{ij}^{\alpha\beta} \left(\bar{\cal O}_{\alpha n}^{C_i} a_{ni}^\dagger + H.c.\right)
  \left(\bar{\cal O}_{\beta n'}^{C_j} a_{n'j}^\dagger + H.c.\right),
 \end{align}
where $\bar{\cal O}_{\alpha n}^{C_i}=\bras{n;C_i}\delta{\cal O}_{\alpha i}\kets{0;C_i}$, which depends only on the sublattice to which the site $i$ belongs.
Moreover, ${\cal H}^{\rm MF}$ is also represented by the bosonic operators as
\begin{align}
  {\cal H}^{\rm MF}=\sum_{in} \Delta E_n^{C_i} \kets{n;C_i}\bras{n;C_i}
  =\sum_{in} \Delta E_n^{C_i} a_{in}^\dagger  a_{in},
\end{align}
where $\Delta E_n^{C_i}$ is the energy difference between the excited state $\kets{n;C_i}$ and ground state $\kets{0;C_i}$, and the constant terms are omitted.
Therefore, the total Hamiltonian ${\cal H}={\cal H}^{\rm MF}+{\cal H}'$ is approximately given by a bilinear form of the bosonic operators as ${\cal H}_{\rm SW}$, which is the spin-wave (SW) Hamiltonian.

The site $i$ is identified by the two indices for unit cell, $l$, and for sublattice $C$.
Here, we introduce the new label $s=(C,n)$ for the two indices $C$ for sublattice and $n$ for the local excited states.
The labels $(in)$ is represented as $(ls)$ for the bosonic operators.
Then, the SW Hamiltonian is generally given as
\begin{align}
{\cal H}_{\rm SW}&=\sum_{ls l' s'}
M_{(ls) (l' s')}^{11}a_{ls}^\dagger a_{l's'}
+\sum_{(ls)< (l' s')}
M_{(ls) (l' s')}^{12}a_{ls}^\dagger a_{l's'}^\dagger
+\sum_{(ls)< (l' s')}
M_{(ls) (l' s')}^{21}a_{ls} a_{l's'}
\nonumber\\
&=\frac{1}{2}\sum_{ls l' s'}\left[
M_{(ls) (l' s')}^{11} a_{ls}^\dagger a_{l's'}
+ M_{(ls) (l' s')}^{11} \left( a_{l's'} a_{ls}^\dagger -\delta_{ll'}\delta_{ss'}\right)
+M_{(ls) (l' s')}^{12}a_{ls}^\dagger a_{l's'}^\dagger +{\rm H.c.}
\right],
\end{align}
where $M^{11}$ and $M^{22}$ are Hermitian matrices satisfying $M^{22}=(M^{11})^T$, $M^{12}$ is a symmetric matrix, and $M^{21}=(M^{12})^*=(M^{12})^\dagger$.
We introduce the operator $a_{lsp}^\dagger$ ($p=1,2$) with $a_{ls1}^\dagger = a_{ls}^\dagger$ and $a_{ls2}^\dagger = a_{ls}$.
The SW Hamiltonian is rewritten as
\begin{align}
  {\cal H}_{\rm SW}=\frac{1}{2}\sum_{ll'}
  {\cal A}_{l}^\dagger M_{l l'} {\cal A}_{l}-\frac{1}{2}\sum_{l}{\rm Tr}M_{ll}^{11},\label{eq:supp-HswAl}
 \end{align}
where we define the vector ${\cal A}_{l}^\dagger$ with respect to the indices $(sp)$ as $[{\cal A}_{l}^\dagger]_{sp} =a_{lsp}^\dagger$.

The Fourier transformations of the operators $\{a_{ls}^\dagger, a_{ls}\}$ are introduced as
\begin{align}
  a_{is}=\sqrt{\frac{N_s}{N}}\sum_{\bm{q}}a_{\bm{q}s}e^{i\bm{q}\cdot\bm{r}_{ls}},
  \qquad
  a_{is}^\dagger=\sqrt{\frac{N_s}{N}}\sum_{\bm{q}}a_{\bm{q}s}^\dagger e^{-i\bm{q}\cdot\bm{r}_{ls}},
\end{align}
where the position of the site $i$ is written as $\bm{r}_i=\bm{r}_{ls}$, and $N_s$ is the number of the sublattices, namely, the number of sites in a unit cell.
The summation for $\bm{q}$ is taken for the first Brillouin zone.
The Fourier transformation of ${\cal A}_{l}^\dagger$ is given by
\begin{align}
  {\cal A}_{l}^\dagger
  =\sqrt{\frac{N_s}{N}}\sum_{\bm{q}}{\cal A}_{\bm{q}}^\dagger e^{-i\bm{q}\cdot\bm{r}_{ls}}.\label{eq:supp-Al}
 \end{align}
Here, we find $[{\cal A}_{\bm{q}}^\dagger]_{s1}=a_{\bm{q}s}^\dagger$ and $[{\cal A}_{\bm{q}}^\dagger]_{s2}=a_{-\bm{q}s}$, which are written as
\begin{align}
  {\cal A}_{\bm{q}}^\dagger=
  \begin{pmatrix}
  \bm{a}_{\bm{q}}^\dagger  &  \bm{a}_{-\bm{q}}
  \end{pmatrix},
\end{align}
where $\bm{a}_{\bm{q}}^\dagger$ is the vector with respect to $s$.
By substituting Eq.~\eqref{eq:supp-Al} to Eq~\eqref{eq:supp-HswAl}, we obtain
\begin{align}
  {\cal H}_{\rm SW}=\frac{1}{2}\sum_{\bm{q}}
{\cal A}_{\bm{q}}^\dagger M_{\bm{q}} {\cal A}_{\bm{q}}-\frac{1}{4}\sum_{\bm{q}}{\rm Tr}M_{\bm{q}},\label{eq:supp-HSWAq}
\end{align}
where $M_{\bm{q}}$ is the Hermitian matrix with respect to $(sp)$ as
\begin{align}
  M_{\bm{q}(sp)(s'p')} =M_{\bm{q}(ss')}^{pp'} =\sum_{l-l'}e^{-i\bm{q}\cdot\left(\bm{r}_{ls}-\bm{r}_{l's'}\right)}
  M_{(l-l', 0) (ss')}^{pp'}.
\end{align}
Here, the submatrices $M^{pp'}$ satisfy the following relations:
\begin{align}
  M_{\bm{q}}^{11}=(M_{\bm{q}}^{11})^\dagger,\quad
  M_{\bm{q}}^{22}=(M_{-\bm{q}}^{11})^T,\quad 
  M_{\bm{q}}^{12}=(M_{-\bm{q}}^{12})^T,\quad 
  M_{\bm{q}}^{21}=(M_{-\bm{q}}^{12})^*= (M_{\bm{q}}^{12})^\dagger.\label{eq:supp-Hsubk}
\end{align}
These are simply written as
\begin{align}
  M_{\bm{q}}^{\bar{p}\bar{p}'}=(M_{-\bm{q}}^{p'p})^T,
\end{align}
where $\bar{p}$ is the counterpart of $p$ ($\bar{1}=2$ and $\bar{2}=1$).

\subsection{Bogoliubov transformation}

In this section, we introduce the Bogoliubov transformation shown in Ref.~\cite{COLPA1978327} to diagonalize the SW Hamiltonian given in Eq~\eqref{eq:supp-HSWAq}.
As shown in Eq.~\eqref{eq:supp-Hsubk}, $M_{\bm{q}}$ is divided to the four submatrices with respect to index $p$.
The matrices and vectors introduced in this section are assumed to be divided in the same manner.

Here, we introduce the para-unitary matrix ${\cal T}_{\bm{q}}$ so as to satisfy 
${\cal T}_{\bm{q}}{\cal I}{\cal T}_{\bm{q}}^\dagger  = {\cal T}_{\bm{q}}^\dagger {\cal I}{\cal T}_{\bm{q}} = {\cal I}$, where ${\cal I}={\rm diag}\{1\cdots1,-1\cdots -1\}$ is the para-unit matrix.
Using the para-unitary matrix, $M_{\bm{q}}$ is diagonalized as
\begin{align}
 {\cal T}_{\bm{q}}^\dagger M_{\bm{q}}{\cal T}_{\bm{q}}= \Omega_{\bm{q}}
 ={\rm diag}[\{\omega_{\bm{q}\eta}^1\},\{\omega_{\bm{q}\eta}^2\}],
\end{align}
where $\omega_{\bm{q}\eta}^2=\omega_{-\bm{q}\eta}^1$, which are positive, and $\eta=1,\cdots, N_s{\cal N}$.
The para-unitary matrix is represented by the submatrix $U_{\bm{q}}$ and $V_{\bm{q}}$ as
\begin{align}
{\cal T}_{\bm{q}}=
  \begin{pmatrix}
  U_{\bm{q}} & V_{-\bm{q}}^*\\
  V_{\bm{q}} & U_{-\bm{q}}^*
\end{pmatrix}, \qquad
{\cal T}_{\bm{q}}^{-1}=
  \begin{pmatrix}
  U_{\bm{q}}^\dagger & -V_{\bm{q}}^\dagger\\
  -V_{-\bm{q}}^T & U_{-\bm{q}}^T
\end{pmatrix},
\end{align}
which are obtained from $M_{\bm{q}}$~\cite{COLPA1978327}.

We also introduce a vector of the operators as
\begin{align}
{\cal B}_{\bm{q}}=
\begin{pmatrix}
  \bm{\alpha}_{\bm{q}}\\
  \bm{\alpha}_{-\bm{q}}^\dagger
\end{pmatrix}
={\cal T}_{\bm{q}}^{-1}{\cal A}_{\bm{q}}=
{\cal T}_{\bm{q}}^{-1}
\begin{pmatrix}
  \bm{a}_{\bm{q}}\\
  \bm{a}_{-\bm{q}}^\dagger
\end{pmatrix},
\end{align}
where $\bm{\alpha}_{\bm{q}}$ is a bosonic operator, which is a so-called Bogoliubov boson.
Using the Bogoliubov bosons, the Hamiltonian shown in Eq.~\eqref{eq:supp-HSWAq} is rewritten as
\begin{align}
  {\cal H}_{\rm SW}&=\frac{1}{2}\sum_{\bm{q}}
{\cal B}_{\bm{q}}^\dagger \Omega_{\bm{q}} {\cal B}_{\bm{q}}-\frac{1}{4}\sum_{\bm{q}}{\rm Tr}M_{\bm{q}}\nonumber\\
&=\sum_{\bm{q}\eta}
  \omega_{\bm{q}\eta} \alpha_{\bm{q}\eta}^\dagger \alpha_{\bm{q}\eta}
  +\frac{1}{4}\sum_{\bm{q}}{\rm Tr}\Omega_{\bm{q}}
  -\frac{1}{4}\sum_{\bm{q}}{\rm Tr}M_{\bm{q}},\label{eq:supp-HSWBq}
\end{align}
where $\omega_{\bm{q}\eta}=\omega_{\bm{q}\eta}^1=\omega_{-\bm{q}\eta}^2$.
The vacuum of the Bogoliubov bosons is defined as $\ketsd{0}$ such that $\alpha_{\bm{q}\eta}\ketsd{0}=0$.

\section{Dynamical correlation function}

In this section, we introduce the formalism to calculate the dynamical correlation function using the SW theory.
The dynamical correlator between 
$\delta{\cal O}_{\alpha i}$ given in Eq.~\eqref{eq:supp-Op} are defined by
\begin{align}
  {\cal S}^{\alpha\alpha'}(\bm{q},\omega)
  &=\frac{1}{2\pi}\int_{-\infty}^\infty dt \brasd{0}\delta {\cal O}_{\alpha\bm{q}}(t) \delta {\cal O}_{\alpha'-\bm{q}}\ketsd{0}e^{i\omega t},
\end{align}
where $\delta {\cal O}_{\alpha \bm{q}}=N^{-1/2}\sum_{i}\delta {\cal O}_{\alpha i} e^{-i\bm{q}\cdot \bm{r}_i}$ and
${\cal O}(t)=e^{i{\cal H}_{\rm SW}t}{\cal O}e^{-i{\cal H}_{\rm SW}t}$. 
From Eq.~\ref{eq:supp-Oi}, $\delta {\cal O}_{\alpha \bm{q}}$ is written using the SW approximation as
\begin{align}
  \delta {\cal O}_{\alpha \bm{q}}\simeq 
  N^{-1/2}\sum_{i} e^{-i\bm{q}\cdot \bm{r}_i}\left( \bar{\cal O}_{\alpha n}^{C_i} a_{ni}^\dagger + H.c.\right)
  =N_s^{-1/2} \sum_{s}\bar{\cal O}_{\alpha s} a_{-\bm{q}s}^\dagger + H.c.,
\end{align}
where we introduce the index $s=(C,n)$ for identifying the sublattice $C$ and local excited state $n$, and $\bar{\cal O}_{\alpha n}^C$ is simply written as $\bar{\cal O}_{\alpha s}$.
Then, the dynamical correlator is calculated as
\begin{align}
  {\cal S}^{\alpha\alpha'}({\bm q}, \omega)&=\sum_{\bm{q}'\eta}\brasd{0}\delta{\cal O}_{\alpha\bm{q}}\ketsd{\bm{q}'\eta}
  \brasd{\bm{q}'\eta}\delta{\cal O}_{\alpha'-\bm{q}}\ketsd{0}\delta(\omega-\omega_{\bm{q}'\eta})\nonumber\\
  &=\frac{1}{N_s}\sum_{\eta} \tilde{W}_{\alpha\bm{q}\eta} \tilde{W}_{\alpha'\bm{q}\eta}^* \delta(\omega-\omega_{\bm{q}\eta}),
\end{align}
where we define the one-boson state as $\ketsd{\bm{q}\eta}=\alpha_{\eta\bm{q}}^\dagger\ketsd{0}$, and
\begin{align}
\tilde{W}_{\alpha\bm{q}\eta}=\sum_{s}\left(
\bar{\cal O}_{\alpha s} V_{\bm{q}s\eta}+\bar{\cal O}_{\alpha s}^{*}U_{\bm{q}s\eta}
\right).
\end{align}

In numerical calculations, we need to introduce the broadening factor $\delta$ as
\begin{align}
  {\cal S}^{\alpha\alpha'}(\bm{q},\omega)
  &=\frac{1}{2\pi}\int_{-\infty}^\infty dt \brasd{0}\delta {\cal O}_{\alpha\bm{q}}(t) \delta {\cal O}_{\alpha'-\bm{q}}\ketsd{0}e^{i\omega t-\delta |t|}\nonumber\\
  &=\frac{1}{N_s}\sum_{\eta} \tilde{W}_{\alpha\bm{q}\eta} \tilde{W}_{\alpha'\bm{q}\eta}^* g(\omega-\omega_{\bm{q}\eta}; \delta),
\end{align}
where $g(\omega; \delta)=\frac{1}{\pi}\frac{\delta}{\omega^2+\delta^2}$ is the Lorentz distribution function.

\section{Transport coefficient}

\subsection{Thermal current}

In this section, we introduce the thermal current.
The energy polarization is defined as
\begin{align}
 \bm{P}_E=\sum_{ls}\bm{r}_{ls}h_{ls},
\end{align}
where $h_{ls}$ is the part of the SW Hamiltonian involving the site $(ls)$, which is represented as
\begin{align}
 h_{ls}=\frac{1}{2}\sum_{p l' s'p'}
 a_{lsp}^\dagger M_{(ls) (l' s')}^{pp'} a_{l's'p'}.
\end{align}
Using this, the SW Hamiltonian given in Eq.~\eqref{eq:supp-HswAl} is written as
\begin{align}
 {\cal H}_{\rm SW}=\sum_{ls}h{_{ls}}.
\end{align}
Here, we define the energy current as 
\begin{align}
 \bm{J}_E=\frac{\partial \bm{P}_E}{\partial t}=i[{\cal H}_{\rm SW},\bm{P}_E].\label{eq:supp-JEcom}
\end{align}
This is equivalent to the thermal current in the bosonic system with the chemical potential $\mu=0$.
By evaluating the commutation relation in Eq.~\eqref{eq:supp-JEcom}, we obtain
\begin{align}
  \bm{J}_E=\frac{1}{2}\sum_{\bm{q}}
  {\cal B}_{\bm{q}}^\dagger
  \bm{{\cal E}}_{\bm{q}}
  {\cal B}_{\bm{q}},
\end{align}
where $\bm{{\cal E}}_{\bm{q}}$ is defined as
\begin{align}
  \bm{{\cal E}}_{\bm{q}}=\frac{1}{2}\left(\bm{{\cal V}}_{\bm{q}}{\cal I}\Omega_{\bm{q}}
  +\Omega_{\bm{q}}{\cal I}\bm{{\cal V}}_{\bm{q}}\right),\label{eq:supp-Edef}
\end{align}
and velocity $\bm{{\cal V}}_{\bm{q}}$ is introduced as
\begin{align}
  \bm{{\cal V}}_{\bm{q}}={\cal T}_{\bm{q}}^\dagger \frac{\partial M_{\bm{q}}}{\partial \bm{q}} {\cal T}_{\bm{q}}.\label{eq:supp-Vdef}
\end{align}
Note that $\bm{{\cal V}}_{\bm{q}}$ and $\bm{{\cal E}}_{\bm{q}}$ are Hermitian matrices and satisfy
$\bm{{\cal V}}_{\bm{q}}^{pp'}=-(\bm{{\cal V}}_{-\bm{q}}^{\bar{p}'\bar{p}})^T$ and $\bm{{\cal E}}_{\bm{q}}^{pp'}=\left(\bm{{\cal E}}_{-\bm{q}}^{\bar{p}'\bar{p}}\right)^T$, respectively.
We also find that Eq.~\eqref{eq:supp-Edef} is rewritten as
\begin{align}
  \bm{{\cal E}}_{\bm{q}\eta\eta'}^{11}=\frac{1}{2}\left(\omega_{\bm{q}\eta}
    +\omega_{\bm{q}\eta'}\right)\bm{{\cal V}}_{\bm{q}\eta\eta'}^{11},\quad
  \bm{{\cal E}}_{\bm{q}\eta\eta'}^{12}=\frac{1}{2}\left(\omega_{\bm{q}\eta}-\omega_{-\bm{q}\eta'}\right)\bm{{\cal V}}_{\bm{q}\eta\eta'}^{12},
\end{align}
which are given in Refs.~\cite{Matsumoto2011,Matsumoto2014}.

\subsection{Current of physical quantities}

In this section, the current of physical quantities such as spin is introduced.
We consider the local quantity ${\cal O}_{ i}$.
To define the current, the total, ${\cal O}_{\rm tot} =\sum_i{\cal O}_{ i}$, should be conserved, namely,
\begin{align}
  \left[{\cal H}_{\rm SW}, {\cal O}_{\rm tot}\right]=0.\label{eq:supp-consO}
\end{align}
In addition, ${\cal O}_{\rm tot}$ should commute also with ${\cal H}^{\rm MF}$, and therefore, 
\begin{align}
  \left[{\cal H}_i^{\rm MF}, {\cal O}_{ i}\right]=0.
\end{align}
This indicates that ${\cal O}_{ i}$ is diagonal on the basis of the eigenstates of ${\cal H}_i^{\rm MF}$.
Thus, ${\cal O}_{\rm tot}$ and its polarization operator for ${\cal O}_{ i}$ are written as
\begin{align}
  {\cal O}_{\rm tot}&=\sum_{ls}\Delta O_{ s} a_{ls}^\dagger a_{ls} + {\rm const.},\label{eq:supp-Odef}\\
  \bm{P}_{{\cal O}}&=\sum_{ls}\bm{r}_{ls}\Delta O_{ s} a_{ls}^\dagger a_{ls} + {\rm const.},
\end{align}
where we use the Holstein-Primakoff transformation given in Eq.~\eqref{eq:supp-HP}, and $\Delta O_{s}$ is defined by
\begin{align}
  \Delta O_{ s}=\bras{n;C}{\cal O}_{ C}\kets{n;C}-\bras{0;C}{\cal O}_{ C}\kets{0;C}.
\end{align}
Here, we use the fact that $\bras{m;C_i}{\cal O}_{i}\kets{m;C_i}$ ($m=0,1,\cdots {\cal N}$) is not dependent on the index for unit cell, $l$, and hence, it is written as $\bras{n;C}{\cal O}_{C}\kets{n;C}$.
Then, we calculate the current as
\begin{align}
  \bm{J}_{{\cal O}}&=\frac{\partial \bm{P}_{{\cal O}}}{\partial t}=i[{\cal H},\bm{P}_{{\cal O}}]\nonumber\\
&=\frac{1}{2}\sum_{\bm{q}}
{\cal B}_{\bm{q}}^\dagger
\bm{{\cal O}}_{\bm{q}}
{\cal B}_{\bm{q}},\label{eq:suppl-Jodef}
\end{align}
where we use Eq.~\eqref{eq:supp-consO}, and $\bm{{\cal O}}_{\bm{q}}$ is defined as
\begin{align}
  \bm{{\cal O}}_{\bm{q}} =\frac{1}{2}
  \left(
    \bm{{\cal V}}_{\bm{q}}{\cal I}\tilde{O}_{\bm{q}}
    +\tilde{O}_{\bm{q}}{\cal I}\bm{{\cal V}}_{\bm{q}}\right).\label{eq:supp-Oqdef}
\end{align}
Here, we introduce
$\tilde{O}_{\bm{q}}={\cal T}_{\bm{q}}^\dagger \tilde{O} {\cal T}_{\bm{q}}$, where $\tilde{O}$ is the diagonal matrix with $[\tilde{O}]_{ss'}^{pp'}=\Delta O_{s}\delta_{ss'}\delta_{pp'}$.
Note that $\bm{{\cal O}}_{\bm{q}}$ is the Hermitian matrix satisfying $\bm{{\cal O}}_{\bm{q}}^{pp'}=\left(\bm{{\cal O}}_{-\bm{q}}^{\bar{p}'\bar{p}}\right)^T$.

\subsection{Kubo formula}

In this section, we show transport coefficients calculated from the Kubo formula.
Given that the current $\bm{J}_{{\cal O}}$ of the quantity ${\cal O}_{\rm tot}$ is caused by the thermal gradient ${\nabla T}$, the expectation value of the current in the presence of the thermal gradient, $\means{J^\mu_{\cal O}}_{\nabla T}$, is represented as 
\begin{align}
 \frac{\means{J^\mu_{{\cal O}}}_{\nabla T}}{V}=L_{{\cal O}}^{\mu\nu}\left(-\frac{\nabla_\nu T}{T}\right),
\end{align}
where $V$ is the volume of the system and $\mu$ and $\nu$ stand for the labels of the Cartesian coordinate.
The coefficient $L_{{\cal O}}^{\mu\nu}$ is calculated by the Kubo formula as
\begin{align}
  L_{{\cal O}}^{\mu\nu}&=\frac{1}{V}\int_0^\infty dt e^{-\delta t}\int_0^{1/T} d\lambda\mean{J_E^\nu(-i\lambda)J_{{\cal O}}^\mu(t)},\label{eq:supp-Lo}
\end{align}
where $\delta$ is a positive infinitesimal constant.
and the conductivity is given by
\begin{align}
 \kappa_{{\cal O}}^{\mu\nu}=-\frac{1}{V}\frac{\means{J_{{\cal O}}^\mu}_{\nabla T}}{\nabla_\nu T}=\frac{L_{{\cal O}}^{\mu\nu}}{T}.
\end{align}
In the present study, we address the longitudinal component with $\mu=\nu$ but it is known that additional contributions are needed to calculate the transverse component~\cite{Matsumoto2011,Matsumoto2014}.
For example, the longitudinal thermal conductivity and spin conductivity for $S^Z$ are given by $\kappa=\kappa_{E}^{xx}$ and $\kappa_s=\kappa_{S^Z}^{xx}$, respectively.

Using the Wick's theorem, Eq.~\eqref{eq:supp-Lo} is evaluated as
\begin{align}
L_{{\cal O}}^{\mu\nu}
&=
\frac{1}{V}
\sum_{\bm{q}\eta\eta'}
\Biggl[
\frac{1}{i}\frac{n(\omega_{\bm{q}\eta})-n(\omega_{\bm{q}\eta'})}{\omega_{\bm{q}\eta}-\omega_{\bm{q}\eta'}}
\frac{[{\cal O}_{\bm{q}}^{\mu}]_{\eta\eta'}^{11}
[{\cal E}_{\bm{q}}^{\nu}]_{\eta'\eta}^{11}}
{\omega_{\bm{q}\eta}-\omega_{\bm{q}\eta'}+i\delta}
-
\frac{n(\omega_{\bm{q}\eta})-n(-\omega_{-\bm{q}\eta'})}{\omega_{\bm{q}\eta}+\omega_{-\bm{q}\eta'}}
{\rm Im}
\left(
\frac{[{\cal O}_{\bm{q}}^{\mu}]_{\eta\eta'}^{12}
[{\cal E}_{\bm{q}}^{\nu}]_{\eta'\eta}^{21}}
{\omega_{\bm{q}\eta}+\omega_{-\bm{q}\eta'}+i\delta}
\right)
\Biggr],\label{eq:supp-Lo2}
\end{align}
where $n(\omega)=\frac{1}{e^{\omega/T}-1}$ is the Bose distribution function with zero chemical potential.
In the limit of $\delta\to +0$, the conductivity is simply written as
\begin{align}
\kappa_{{\cal O}}^{\mu\nu}=
\frac{\pi}{VT^2}
\sum_{\bm{q}\eta\eta'}n(\omega_{\bm{q}\eta})[1+n(\omega_{\bm{q}\eta})]
{\rm Re}\left[[{\cal O}_{\bm{q}}^{\mu}]_{\eta\eta'}^{11}
[{\cal E}_{\bm{q}}^{\nu}]_{\eta'\eta}^{11}\right]\delta(\omega_{\bm{q}\eta}-\omega_{\bm{q}\eta'}).
\end{align}
Here, we use the fact that the second term of Eq.~\eqref{eq:supp-Lo2} does not contribute because $\omega_{\bm{q}\eta}$ is positive.
If there is no degeneracy in substantial regions of the Brillouin zone, this is rewitten as
\begin{align}
  \kappa_{{\cal O}}^{\mu\nu}\simeq\frac{1}{\delta}
  \frac{1}{VT^2}
  \sum_{\bm{q}\eta}n(\omega_{\bm{q}\eta})[1+n(\omega_{\bm{q}\eta})]
  [{\cal O}_{\bm{q}}^{\mu}]_{\eta\eta}^{11}
  [{\cal E}_{\bm{q}}^{\nu}]_{\eta\eta}^{11}.
  \end{align}
In Eq.~\eqref{eq:supp-Edef}, $\Omega_{\bm{q}}$ is diagonal, and therefore, $\bm{{\cal E}}_{\bm{q}\eta\eta}^{11}$ is rewritten as
\begin{align}
  \bm{{\cal E}}_{\bm{q}\eta\eta}^{11} =
  \bm{{\cal V}}_{\bm{q}\eta\eta}^{11}\omega_{\bm{q}\eta}
  =\bm{v}_{\bm{q}\eta}\omega_{\bm{q}\eta},
\end{align}
where we use the following relation
\begin{align}
  \bm{{\cal V}}_{\bm{q}\eta\eta}^{11}=\frac{\partial \omega_{\bm{q}\eta}}{\partial \bm{q}}
  \equiv \bm{v}_{\bm{q}\eta}.
\end{align}
In the numerical calculations, it is difficult to evaluate the derivative $\frac{\partial \omega_{\bm{q}\eta}}{\partial \bm{q}}$ numerically.
Instead, $\bm{{\cal V}}_{\bm{q}\eta\eta}$ is computed using Eq.~\eqref{eq:supp-Vdef}.
We also find from Eqs.~\eqref{eq:supp-consO} and \eqref{eq:supp-Odef} that 
the transformation matrix ${\cal T}_{\bm{q}}$ can be chosen so that
$\tilde{O}_{\bm{q}}$ is diagonal in Eq.~\eqref{eq:supp-Oqdef}, which leads to $\bm{{\cal O}}_{\bm{q}\eta\eta}^{11} =\bm{v}_{\bm{q}\eta}\tilde{O}_{\bm{q}\eta\eta}^{11}$.
Thus, the conductivity is represented as
\begin{align}
  \kappa_{O}^{\mu\nu}
\simeq \frac{1}{\delta}
\frac{1}{VT^2}
  \sum_{\bm{q}\eta}n(\omega_{\bm{q}\eta})[1+n(\omega_{\bm{q}\eta})]
  v^{\mu}_{\bm{q}\eta}v^{\nu}_{\bm{q}\eta}\omega_{\bm{q}\eta}\tilde{O}_{\bm{q}\eta\eta}^{11}.\label{eq:supp-kappa-kubo}
\end{align}

\subsection{Boltzmann equation}

In this section, we show the form of the conductivity obtained by the Boltzmann equation.
Since $\tilde{O}_{\bm{q}}$ defined in Eq.~\eqref{eq:supp-Oqdef} is a diagonal matrix, the current operator in Eq.~\eqref{eq:suppl-Jodef} is written as
\begin{align}
  \bm{J}_O&=\frac{1}{2}\sum_{\bm{q}\eta}
  \left(
  \bm{{\cal O}}_{\bm{q}\eta\eta}^{11}
  \alpha_{\bm{q}\eta}^\dagger\alpha_{\bm{q}\eta}
+  \bm{{\cal O}}_{-\bm{q}\eta\eta}^{11}
\alpha_{-\bm{q}\eta}\alpha_{-\bm{q}\eta}^\dagger
  \right)\nonumber\\
  &=\sum_{\bm{q}\eta}
  \bm{{\cal O}}_{\bm{q}\eta\eta}^{11}
  \alpha_{\bm{q}\eta}^\dagger\alpha_{\bm{q}\eta},
\end{align}
where we use $\sum_{\bm{q}\eta}\bm{{\cal O}}_{\bm{q}\eta\eta}^{11}=0$.
In the equilibrium system without a thermal gradient, the current should vanish, and hence, the expectation value in the presence of the thermal gradient $\nabla_{\nu} T$ is given by
\begin{align}
  \means{J_O^{\mu}}_{\nabla_{\nu} T}=\sum_{\bm{q}\eta}
  [{\cal O}_{\bm{q}}^\mu]_{\eta\eta}^{11}
  \left(n_{\bm{q}\eta}^{\nabla_{\nu} T}(\bm{r})
  -n_{\bm{q}\eta}\right),\label{eq:supp-Jo-boltz}
\end{align}
where $n(\omega_{\bm{q}\eta})=n_{\bm{q}\eta}$ and $n_{\bm{q}\eta}^{\nabla_{\nu} T}(\bm{r})$ is the spatial-dependent Bose distribution function of the steady state in the presence of the thermal gradient $\nabla_{\nu} T$.
By applying the relaxation time approximation to the Boltzmann equation, we obtain 
\begin{align}
  n_{\bm{q}\eta}^{\nabla_{\nu} T}(\bm{r})
-n_{\bm{q}\eta}\simeq -\tau_{\bm{q}\eta} v_{\bm{q}\eta}^{\nu} \nabla_{\nu} n_{\bm{q}\eta}
=-\tau_{\bm{q}\eta} v_{\bm{q}\eta}^{\nu} \frac{\partial n_{\bm{q}\eta}}{\partial T}\nabla_{\nu} T,
\end{align}
where $\tau_{\bm{q}\eta}$ is the relaxation time for the quasiparticle with $(\bm{q},\eta)$.
By substituting this in Eq.~\eqref{eq:supp-Jo-boltz}, we find
\begin{align}
  \means{J_O^{\mu}}_{\nabla_{\nu} T}
  &\simeq 
  -\sum_{\bm{q}\eta}
  [{\cal O}_{\bm{q}}^\mu]_{\eta\eta}^{11}
  \tau_{\bm{q}\eta} v_{\bm{q}\eta}^{\nu} \frac{\partial n_{\bm{q}\eta}}{\partial T}\nabla_{\nu} T
  =-\sum_{\bm{q}\eta}
  v^{\mu}_{\bm{q}\eta}\tilde{O}_{\bm{q}\eta\eta}^{11}
  \tau_{\bm{q}\eta} v_{\bm{q}\eta}^{\nu} \frac{\partial n_{\bm{q}\eta}}{\partial T}\nabla_{\nu} T\nonumber\\
  &=
\frac{1}{T^2}
  \sum_{\bm{q}\eta}
  \tau_{\bm{q}\eta}
  n(\omega_{\bm{q}\eta})[1+n(\omega_{\bm{q}\eta})]
  v^{\mu}_{\bm{q}\eta}v^{\nu}_{\bm{q}\eta}\omega_{\bm{q}\eta}\tilde{O}_{\bm{q}\eta\eta}^{11}(-\nabla_{\nu} T).
\end{align}
When the dependence on $(\bm{q},\eta)$ of $\tau_{\bm{q}\eta}$ is neglected, 
the thermal conductivity is written as
\begin{align}
  \kappa_{O}^{\mu\nu}
= 
\frac{\tau}{VT^2}
  \sum_{\bm{q}\eta}n(\omega_{\bm{q}\eta})[1+n(\omega_{\bm{q}\eta})]
  v^{\mu}_{\bm{q}\eta}v^{\nu}_{\bm{q}\eta}\omega_{\bm{q}\eta}\tilde{O}_{\bm{q}\eta\eta}^{11}.
\end{align}
This coincides with Eq.~\eqref{eq:supp-kappa-kubo}, where $\delta=1/\tau$ is supposed.

\end{document}